\documentclass[aps,pra,twocolumn,groupedaddress,showpacs]{revtex4}
\usepackage[dvips]{graphics,epsfig}
\usepackage{latexsym}

\begin{document}


\preprint{LBNL-49088}

\title{Improved alternating gradient transport and focusing of neutral
molecules}

\author{Juris Kalnins}
\email[E-mail]{JGKalnins@lbl.gov}
\author{Glen Lambertson}
\email[E-mail:]{GRLambertson@lbl.gov}
\author{Harvey Gould}
\email[E-mail]{Gould@lbl.gov}
\affiliation{Mail Stop 71-259, Lawrence Berkeley National Laboratory,
University of California, Berkeley CA 94720}



\begin{abstract}
Polar molecules, in strong-field seeking states, can be 
transported and focused by 
an alternating sequence of electric field gradients 
that focus in one transverse direction 
while defocusing in the other. We show, by calculation and 
numerical simulation, how one may greatly improve the 
alternating gradient transport and focusing of molecules. 
We use a new optimized multipole lens design, a FODO-lattice
beam transport line, and lenses to match the beam transport line to
the beam source and to the final focus.

We derive analytic expressions for the potentials, fields, and
gradients that may be used to design these lenses. We describe a simple
lens optimization procedure and derive the equations of motion for 
tracking molecules through a beam transport line. As an example, we model 
a straight beamline that transports a 560 m/s jet-source beam of 
methyl fluoride15 m from its source and focuses it to 2 mm diameter.
We calculate the beam transport line acceptance and beam survival, 
for a beam with a velocity spread, and estimate the transmitted intensity for 
specified source conditions. Possible applications are discussed. 
\end{abstract}



\maketitle

\section {Introduction \label{intro}}
\begin{figure}
\includegraphics{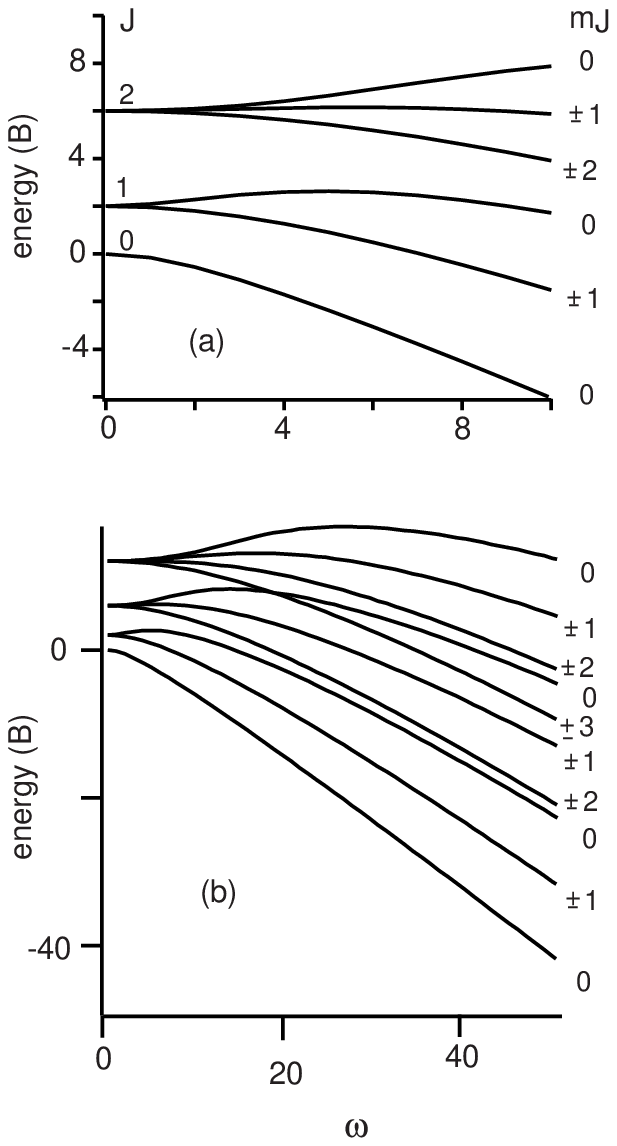}
\caption{\label{stark} Energy levels of low-lying rotational states of a
diatomic or symmetric-top molecule in an electric field (Stark effect).
The levels were  calculated using the rigid rotor model. 
The horizontal axis is in dimensionless units of $\omega = d_e E/B$, 
where $d_e$ is the electric
dipole moment, E the electric field, and B the rotational constant. The
vertical axis is in units of B.  Energies for small $\omega$ are shown in (a)
and energies for large $\omega$ in (b). 
The potential energy of the molecule, $W$, is the change in energy from 
zero electric field and for very large $\omega W \to d_eE$.
The levels that descend are strong-field seeking (see text). 
For the $J = 0$ state of methyl fluoride ($\mathrm {CH_3F}$), 
in the K = 0 state, $\omega = 1$ at 2.78 MV/m. For the $J
= 0$ state of CsF, $\omega = 1$ at 0.135 MV/m. } \end{figure}

A polar molecule has an intrinsic separation of charge 
on which an electric field gradient exerts a force.
The force, $F_x$, in the (transverse) $x$ direction is:
%
\begin{equation} \label{1} F_x = - \frac {\partial W}{\partial x} = - \frac
{\partial W}{\partial E} \frac { \partial E}{\partial x}
\end{equation}
where $W$ is the potential energy of the
molecule in an electric field E (Stark effect) of magnitude $E = (E_x^2 +
E_y^2)^{1/2}$; and similarly for the force, $F_y$ in the $y$ direction.
The force and gradient are in opposite directions (weak-field seeking)
for rotational states that become less tightly bound in an electric field 
($\partial W/ \partial E >0$), while the force and gradient are in the same
direction (strong-field seeking) for rotational states that become 
more tightly bound in an electric field ($\partial W/ \partial E < 0$). 
The $J$ = 0 state is always strong-field seeking and all rotational states
become strong-field seeking in the limit of strong electric field as shown in Fig.~\ref{stark}.
%

Focusing a beam of molecules, traveling in the $z$ direction, is achieved 
using static two-dimensional ($x$, $y$) electric field gradients. We
neglect end field effects and assume $E_z = 0$ inside the focusing lenses.
Molecules in weak-field seeking states can be focused, in both directions, 
using quadrupole and/or sextupole fields that have a minimum in 
the electric field in both directions.
Molecules in strong-field seeking states, however, can be focused in only
one transverse direction, while being defocused in the other, because it is not possible to
have a maximum in the electric field in both dimensions (in free space).

Molecules in strong-field seeking states have been transported and focused
by an alternating sequence of electric field gradient lenses \cite{auerbach66, kakati67, kakati69,
gunther72, lubbert75, lubbert78, reuss88} (as have neutral atoms
\cite{noh00}), but it has been neither as successful, nor as widely used, as
has quadrupole and sextupole focusing for molecules in weak-field seeking states
\cite{reuss88, cho91}. 

In this paper we show, by calculation and numerical simulation, how one may greatly
improve the alternating gradient transport and focusing of molecules by
optimizing the lens field geometries so that nonlinearity in 
$\partial W / \partial E$ is compensated by $\partial E / \partial x$
to produce a linear restoring force.
We use this optimized multipole lens design 
in a FODO-lattice beam transport line and use additional lenses to match the
beam transport line to the beam source and to the final focus. 

The remainder of this paper is organized as follows: Section
\ref{alternate_gradient} discusses alternating gradient focusing, transport lattices,
matching lenses, and linear optics. Section
\ref{linear_optics} derives the formulae for designing linear focusing elements
and presents examples of lenses. Section
\ref{beam_transport} derives the equations for molecular beam transport 
and models both a simple 30 m-long FODO lattice and a complete 
15 m-long transport line with entrance and
exit matching lenses. Section \ref{intensity} estimates the intensity of a
methyl fluoride jet-source beam, transported through a beamline and
focused. Section \ref{uses} discusses the use of strong-field seeking states, and 
possible applications of alternating grading focusing and transport through very long 
beamlines.
\section{\label{alternate_gradient} Alternating gradient beam transport
and focusing} 
A beam of charged particles can be focused and transported over almost
unlimited distances by alternating F and D type magnetic quadrupole
lenses. The F type lens focuses the beam in the horizontal ($x$) direction while
defocusing it in the vertical ($y$) direction. The D type defocuses the beam
in the horizontal direction while focusing it in the vertical  direction.

\begin{figure*}
\includegraphics{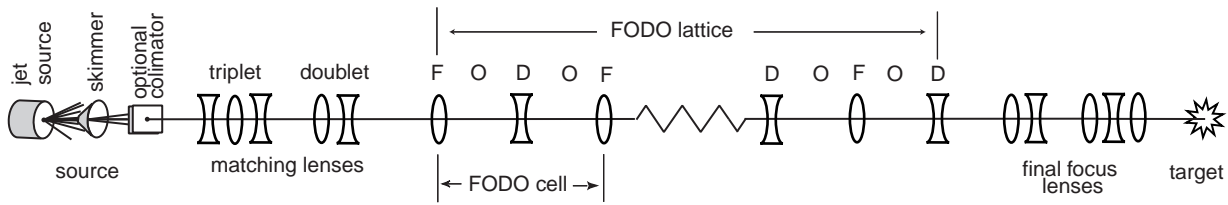}
\caption{\label{beamline} Schematic diagram of an alternating gradient
beam transport line for molecules. It consists of a jet source and skimmer,
beam matching optics, FODO lattice, and final focus optics.}
\end{figure*}

A complete alternating gradient transport line begins with a beam source whose
output is optically matched, by lenses, into the acceptance of a transport
section which in turn is matched, by lenses, to a final focus. 
The final focus can be, for example, at an experimental target or at
a matching point for injection into a (storage) ring lattice. A complete transport line
for molecules, starting with a jet source and skimmer, is shown in Fig.
\ref{beamline}.  %
%
For long-distance beam transport, the lens
system providing the highest beam acceptance and requiring the lowest
focusing strength is a lattice of identical FODO-cells, in which the F/D-
lenses are separated by drift spaces (O).

To match the molecular beam source to the transport line, 
and to match the transport line to the final focus,
we use doublet (FD/DF) and triplet (FDF/DFD) type lens configurations.
These lens configurations generally require higher focusing strengths than do
the FODO cells. Doublets are typically used for asymmetric focal points
(unequal $x$ and $y$ dimensions) while triplets are used for symmetric ones.  
In the F/D
doublet, a net focusing in $x$ and $y$ occurs because particles first
focused in $x$ are, in the second element, closer to the axis and
therefore less defocused. Particles first defocused in $y$ are, in the
second element, further from the axis and thus more strongly focused.

In all lenses, linear focusing is needed for optimum optical properties. Linear focusing
requires that the force in each direction be linearly proportional
to the displacement in that direction ($|F_x /x| = $ constant and $|F_y /y| = $
constant), and independent of the displacement in the other direction
(uncoupled motion). Strong non-linearities in focusing elements will
result in loss of beam, generation of beam halo, and growth in the
transverse emittance (the product of the angular
divergence and the spatial dimension) producing larger beam sizes. 
\section{Linear optics\label {linear_optics}}
\subsection {\label{potential_energy} Potential energy of a molecule in
an electric field} 
To determine the lens shape that will produce the most linear force on the molecule,
we need $\partial W / \partial E$ for Eq.~(\ref{stark}). 
This quantity will change with the
$J, |m_J|$ rotational state and the electric field strength. 
If a number of rotational levels have a similar $\partial W / \partial E$,
then one lens design will be nearly optimum for all of them. 
In the limiting case of a strong electric field (large $\omega$ in Fig.~{\ref{stark} ),
$\partial W / \partial E = -d_e$. 

For polar molecules in weaker fields, we calculate the interaction energy,
in the rigid rotor approximation, following the approach of von Meyenn 
\cite{vonmeyen}. The Hamiltonian operator is $H = B \mathbf J^2 - d_e E
\cos \theta$, where B is the rotational constant,
and the direction cosine matrix elements, which couple ($J$,
$m_J$) with ($J +1$, $m_J$) and ($J-1$, $m_J$), are taken from Townes
and Schawlow \cite{townes55}. We diagonalize the matrix for terms
through $J = 30$. The first few levels are shown in Fig.~\ref{stark}.
As $\omega = d_eE/B$ and $B$ determine the Stark effect 
for each $J, |m_J|$, it is straight forward to construct a simple function for the
Stark effect for any $J, |m_J|$. For the $J = 0$ state of a diatomic molecule or symmetric top (K = 0)
molecule, a satisfactory approximation \cite {schwan00} is:
\begin{equation}\label{stark_shift}
W(E) = \frac {C_1 \omega ^2 B}{1+ C_2 \omega } = \frac {C_1 d_e^2 E^2}{B
+ C_2 d_e E} 
\end{equation}
with $C_1 = -0.2085$  and $C_2 = 0.2445$. This expression works best
for small and intermediate values of $\omega$.

From Eq.~(\ref{stark_shift}), for the $J = 0$ state, the field derivative of the
potential is: 
\begin{equation}\label{slope}
\frac{\partial W} {\partial E}  = \frac {W}{E} \left [\frac {2B +C_2 d_e E}{B
+C_2 d_e E} \right]
\end {equation}
Similar expressions may be found for other rotational levels.
A perturbation expression for W
can be used for rotational states in (very) weak electric fields
\cite{townes55}. In some applications, using $\partial W/ \partial E$ from either the strong field limit
or the weak field limit will be sufficient to design a linear lens.

Finally, we note that all atoms and molecules, including nonpolar
molecules, are polarizable with an interaction energy $W_{\alpha} = - \frac{1}{2}\alpha E^2$,
where $\alpha$ is the dipole polarizability \cite{miller99, miller77}. For
laboratory electric fields this interaction is much smaller than the
interaction with a molecular electric dipole moment, but it can be used to
focus atoms \cite{noh00} or decelerate them \cite{maddi99}.

\begin{widetext}
\subsection{Electric field gradient of a focusing lens}
Any lens electrostatic potential, $\Phi$ can be expressed as the 
following multipole expansion in
cylindrical coordinates, $r$, $\theta$:
\[ -\Phi(r, \theta) = E_0r_0 \left[ \sum_{n=1}^{\infty} \frac{b_n}{n}
\left(\frac{r}{r_0}\right)^n \cos (n \theta) + \sum_{n=1}^{\infty}
\frac{a_n}{n} \left(\frac{r}{r_0}\right)^n  \sin (n \theta ) \right] \]
where $E_0$ is the central field [$E = E_0(b_1^2 + a_1^2)^{1/2}$], $r_0$
is a scaling length, and $b_n$ ($a_n$) are the dimensionless constants of
the $2n$- pole strengths for normal (skew) elements. 

Setting $b_n = a_n$ for simplicity, converting to Cartesian coordinates,
and retaining only the $a_1$, $a_3$, and $a_5$ terms 
(see section \ref{force}), the lens potentials for normal multipoles are: 
\begin{eqnarray}
 - \Phi_N (x,y) = E_0x \left[ a_1 + \frac{a_3 (x^2 -3y^2)}{3 r_0^2} \right. 
\left.+ \frac{a_5 (x^4 - 10x^2y^2 + 5y^4)} {5 r_0^4} \right] \label {normal}
\end {eqnarray}
%
or for skew multipoles are :
\begin {eqnarray} 
- \Phi_S(x,y) = E_0y \left[a_1 + \frac{a_3 (3x^2 - y^2)}{3r_0^2}\right.
\left. + \frac{a_5 (5x^4 -10x^2y^2 + y^4)}{5r_0^4}\right] \label {skew}
\end {eqnarray}
 where $E_0$ is the field on axis for $a_1 = 1$.

Both of these potentials give the same total electric
field ($E = - \nabla \Phi$).  
\begin{eqnarray}
\label{field}
E(x,y) = (E_x^2 + E_y^2)^{1/2} = E_0\left[ a_1^2 + \frac{2a_1a_3 (x^2- y^2)}
{r_0^2}\right.
\left. + \frac{a_3^2 (x^4+2x^2y^2+y^4)}{r_0^4} \right.
\left. + \frac{2a_1a_5 (x^4-6x^2y^2+y^4)}{r_0^4} + \dots \right] ^{1/2}
\label{total_field}
\end{eqnarray}
\end{widetext}
which has the electric field gradients:
\begin{eqnarray}
\frac{\partial E}{ \partial  x } = \frac{2 a_3 E_0^2}{ r_0^2} \frac{G_x x  }{ E}
\nonumber \\
\frac{\partial E}{ \partial y}  = -\frac{2 a_3 E_0^2   }{ r_0^2}\frac{ G_y y}{E}
\label{gradient}
\end{eqnarray} where 
%
\begin{eqnarray}
G_x(x,y) = a_1 + \frac{a_3 (x^2  + y^2)}{r_0^2}
+ \frac{2a_1a_5 (x^2 -3y^2)}{a_3 r_0^2} + \dots\nonumber \\
G_y(x,y) = a_1 - \frac{a_3 (x^2  + y^2)}{r_0^2}
+ \frac{2a_1a_5 (3x^2 -y^2)}{a_3r_0^2}+ \dots   \label{G}
\end{eqnarray}
\subsection{Force on the molecule due to an electric field gradient \label{force}} 
The force on a polar molecule in the $x$ or $y$ direction can now be calculated
using Eq's.~(\ref{1}, \ref{gradient}, and \ref{G}) with a suitable
expression  for $\partial W/ \partial E$ from section \ref{potential_energy}.

Non-linear forces, inside the focusing
lenses, limit the maximum beam size that one can transport without
suffering beam losses, emittance (size) growth, and beam halo. 
Non-linear forces generally arise from higher-order multipole components of the lens
electrostatic potential. Cylindrical electrodes, which are two-wire field lenses (with $r_0 =$ half-gap), 
shown in Fig.~\ref{equipotentials} (a), have long been used to focus molecules in
strong-field seeking states \cite{auerbach66, kakati67, kakati69,
gunther72, lubbert75, lubbert78, reuss88}. They contain multipoles
of all odd orders and of equal strengths ( $a_1 =1$, $a_3 =-1$, $a_5 = 1
\dots$). As we will see below, itÕs strong decapole field ($a_5 =
1$) reduces the area of the lens over which the
focusing is linear (dynamic aperture). 
A beam transport line using two-wire field
lenses will be limited to a smaller diameter beam compared to a similar
transport line using optimized multipole lens such as the one shown in Fig.~\ref{equipotentials} (b).
\begin{figure}
\includegraphics{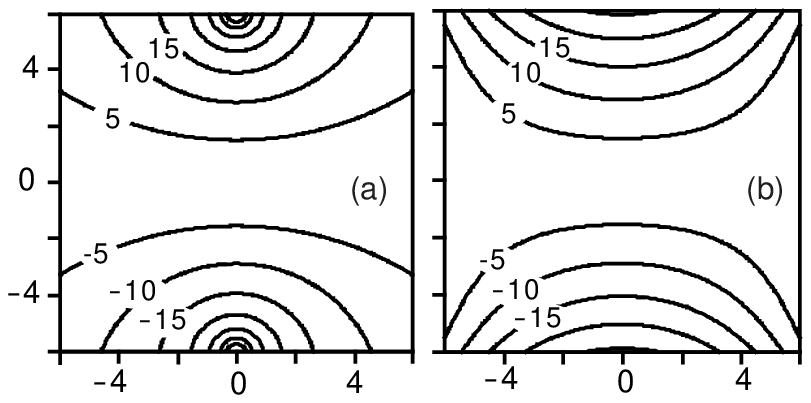}
\caption{\label{equipotentials} Equipotentials (truncated) in $x$ and $y$ of skew F lenses. 
The actual electrode can be fabricated from any set of equipotentials and,
if dimensions are in mm, they match lenses used in the examples in Section \ref{beam_transport}.
A two-wire
field lens (with multipole components: $a_1 =1$, $a_3 =-1$, $a_5 = 1 \dots$) is shown in (a). 
An optimized multipole lens (with multipole strengths $a_1 =1$, $a_3 =-1$, $a_5 = 0$) is shown in (b). 
The lens in (b) is designed to focus in the strong field limit.}
\end {figure}
%

To produce a larger linear region
we try to limit the multipole fields to those that
are essential for the optics. We include a dipole field ($a_1$) to give a non-zero field on
axis, without which strong-field seeking molecules defocus in both
transverse directions; a sextupole field ($a_3$) to provide the linear
focusing force (of order $r^1$); and a decapole field ($a_5$) to correct the non-linear forces (of
order $r^3$) produced by the sextupole field. 
We omit the quadrupole field ($a_2$), because it bends the
beam, and we omit the octupole field ($a_4$), because it introduces stronger
non-linear forces (of order $r^2$) than those of the sextupole/decapole fields. All
other multipole strengths are set to zero. In a real lens, however, the
electrodes may be truncated equipotentials, with the consequence that small
residual higher order multipoles will remain.

Since the lens potentials, as defined by EqÕs.~(\ref{normal}, \ref{skew}),
have two free parameters, we chose $a_1 = 1$, making $E_0$ the
central field and we choose $|a_3| = 1$, making it easier to compare the optimized multipole lens
to a two-wire field lens. The remaining choice is the decapole field strength $a_5$, which we
use to optimize the lens optics. We do this in two ways: first by calculating,
for specific molecules, the forces $F_x$ and $F_y$ inside a lens and comparing their 
linearity in $x$ and $y$ for different $a_5$, and
second, in section \ref{beam_transport}, by simulation of beams in model transport lines.

If we calculate the horizontal force, $F_x$ on a molecule whose potential energy is
$W = -d_eE$ (strong field limit), we find that it is most linear on-axis ($y = 0$) for no
decapole field ($a_5 = 0$). For this example, the constant-force
contours for $F_x$ in the $x, y$ plane are shown, for the two wire field lens, in
Fig.~\ref{Fx} (a) and, for the zero-decapole lens, in Fig.~\ref{Fx} (b). The zero decapole
lens is seen to have far better linearity than the two-wire field lens. 
(This is also true for $F_y$ which is not shown here.) 
In a similar way, we find that the optimized multipole lens outperforms 
the two-wire field lens in the low electric field limit (quadratic Stark effect), where the
most linear field on axis has $a_5 = -\frac{1}{2}$. 
%
%
\begin{figure}
\includegraphics{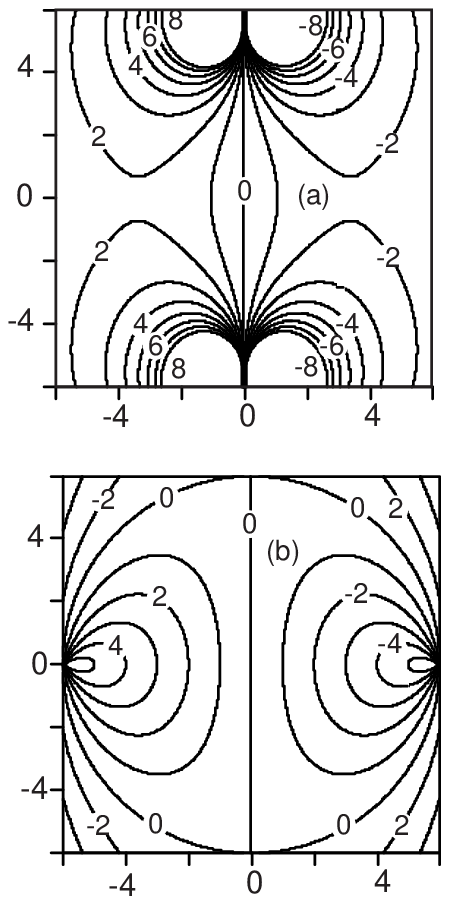}
\caption{\label{Fx} Contour plots of the force in the $x$ direction ($F_x$), on a
molecule inside a lens, as a function of $x$ and $y$ position. The lines are contours
of equal $F_x$ in arbitrary units. In this example we use the 
large Stark effect limit ($W = -d_e E$) and $r_0 = 6$ mm. 
(a) shows $F_x$ for a two-wire field lens. (b) shows $F_x$, for an optimized multipole lens
($a_5$ = 0). A lens that was completely linear in $F_x$ would have vertical contour lines
with uniform spacing.
}
\end {figure}

\subsection{\label{end_effects} End effects}
Real lenses have finite length, and the two-dimensional fields inside become
 three-dimensional fields at the ends. There are three effects from this. 
First, the $z$-component of the field gradient affects the transverse motion. 
This will be small if the lens aperture is small compared to the
physical length of the lens. 

Second, the lens field acts over a distance that is different than the 
physical length of the electrodes
because of (extending) fringe fields at the ends. 
For linear design optics we approximate the real lens by a lens 
of constant central field $E_0$ over an effective length, 
different than the physical length of the lens. 
The effective length is determined by integrating the actual central field 
(found by numerical calculation or measurement) through the lens, 
including the ends, and dividing by $E_0$.

Third, the $z$-component of the electric field gradient, present at the ends of the lens,
has an effect on the $z$-component ( longitudinal) velocity.  
The non-relativistic Hamiltonian (neglecting gravity) for a molecule of
mass $m$ and total velocity $v$ is: $H = \frac {1}{2} m v^2 + W$.
The Hamiltonian is conserved if the electric field is static. Consequently,
a molecule in a strong-field seeking (weak-field seeking) state will gain (lose)
kinetic energy entering the lens and then lose (gain) the same amount of kinetic energy upon
exit \cite{bethlem99, maddi99}. For the central ($x = y = 0$) trajectory,
the longitudinal velocity, $v_{zE}$, of the molecule, in the lens is: 
%
\[ v_{zE}^2  = v_{z0}^2 - W/2 m \]
where $v_{z0}$ is the velocity of the molecule in the
drift spaces. For most thermal or jet-source molecules in laboratory
electric fields, the change in velocity is a small effect. (A beam of 560 m/s
methyl fluoride molecules in the $J = 0$ state increases its velocity by 0.2 m/s
upon entering an electric field of 10 MV/m.)
\section{\label{beam_transport}beam transport}
\subsection{Equations of motion\label{equations}}
To track the trajectories of molecules passing through combinations of lenses,
we use the transverse nonlinear equations of motion 
for a molecule traveling in the $z$ direction, given from 
Eq's.~(\ref{1}, \ref{gradient}, and \ref{G}):
\begin{eqnarray}
 x''  + \frac{2 a_3}{r_0^2} \frac{ E_0^2}{m v_{ZE}^2} \frac{ G_x x}{E} \frac{\partial
W}{\partial E}& =&0 \nonumber \\*
 y'' - \frac{2 a_3}{r_0^2} \frac{ E_0^2}{m v_{ZE}^2} \frac{ G_y y}{E} \frac{\partial
W}{\partial E}& =&0 \label{transverse_motion}
\end{eqnarray}
with trajectory gradients defined as $x' = dx/dz = v_x/v_{ZE}$
and $y' = dy/dz = v_y/v_{ZE}$.

We start the design optics using completely linear lenses ($G_x = G_y = 1$, and $E =
E_0$), for which Eq.~(\ref{transverse_motion}) reduces to:
%
\begin{equation} \label{linear_focus}
 x''  + K_0x = 0   \hskip20pt    y'' - K_0y = 0
\end{equation}
where $K_0$ is the lens linear focusing strength:
\begin{eqnarray} 
K_0 =& \frac {2 a_3}{r_0^2} \frac{ E_0}{ mv_{ZE}^2} \frac{\partial W}{\partial E}&
\nonumber \\*
=& -\frac{2 a_3}{r_0^2} \frac{d_e E_0}{ m v_{ZE}^2} &\text{   for } W = -d_eE
\nonumber \\*
=& -\frac{a_3}{r_0^2} \frac{\alpha E_0^2 }{m v_{ZE}^2} &\text{   for } W_{\alpha} = -
\frac{1}{2} \alpha E^2 
\nonumber \\*
=& \frac{2 a_3}{r_0^2} \frac{ d_e^2 E_0^2 }{ m v_{ZE}^2}
\frac{C_1(2B + C_2d_eE_0)}{(B + C_2d_eE_0)^2} &\text {   for } J = 0 \text {, from Eq.~\ref{stark_shift}}
\label{K0}
\end{eqnarray}

For a molecule in a strong-field seeking state ($\partial
W/ \partial E < 0$), the lens will focus in the $x$ direction (F-lens) for $a_3 <0$,
and focus in the $y$ direction (D-lens) for $a_3 >0$. For a molecule in a 
weak-field seeking state ($\partial W/ \partial E > 0$) the F and D directions
are reversed. The lenses can also be used to focus
atoms ($\partial W_{\alpha}/ \partial E = -\alpha E$)  which, in their ground states, 
are always strong-field seeking. 

The linear design optics will determine the lens focusing strength $K_0$.
Then, the choice of scaling radius $r_0$, will give the required lens field,
$E_0$, from Eq.~(\ref{K0}).
Small $r_0$ allows us to use low electric fields but, as seen from 
Eq's.~(\ref{field}, \ref{G}), large
$r_0$ is needed for increased linearity. The value of $r_0$ chosen will
then depend on the maximum electric field strength and the beam
size to be transported.
\subsection{Example of methyl fluoride\label{methyl_fluoride}}
For a realistic beam transport simulation, we assume a beam of methyl fluoride ($\mathrm
{CH_3F}$), in the $J = 0$ rotational state, having a
longitudinal velocity $v_{Z0} = 560$ m/s. 
(This is the approximate
velocity of a beam produced by seeding methyl fluoride in
an argon jet source with a reservoir temperature of 300 K (See section \ref{intensity}) .
The electric field derivative of the potential
energy (of the $J = 0$ state), is given by
Eq.~(\ref{slope}) with $d_e = 6.25 \times 10^{-30}$ J/V/m (1.86
Debye) and  rotational constant $B = 1.76 \times 10^{-23}$ J ($0.88
\text{ cm}^{-1}$).

To study long distance transport, we model a FODO lattice consisting of a
sequence of identical F and D lenses separated by drift spaces (O). Then, to
study a complete beam line, we add an upstream lens section, to match the
beam from the jet source into the FODO lattice, and a downstream section
for a final focus of the beam (see Fig.~\ref{beamline}).
\subsection {FODO lattice\label{fodo}}
We chose a simple FODO lattice
consisting of identical FODO cells (see Section \ref{alternate_gradient}).
The optics of a FODO cell starts in the
center of an F lens (or D lens) and ends at the center of the next F lens
(or D lens). At the ends of the FODO cell the beam is at a waist (defined as zero slope
in the beam size) with a maximum size in one plane and a minimum size in
the other plane. 

For this study, we use 10 cm-long F/D lenses, with scaling length $r_0 = 6$
mm, separated by 40 cm drift spaces, giving a FODO cell length of 100
cm.  This design leaves 80 \% of the cell unfilled. It is
economical to build, but does not have as large an acceptance
as designs that filled more of the lattice (see section \ref{uses}).

In a FODO lattice the motion of a particle is periodic in
phase-space. The phase-advance measures how far along the
period it has proceeded from its initial starting point.
The transverse linear optics are
characterized by the phase advance in the FODO cell $ \mu_c$. 
For our cell we choose $
\mu_c = \pi /3$ rad, for which all particles return to their initial
phase-space  position after 6 FODO cells.
This specifies a central field of $E_0 = 3.23$ MV/m, which is not close to
breakdown, as well as, a beam size that is nearly the minimum possible.  
\subsection {Modeling beam transmission\label{modeling}}
\subsubsection{beam distribution in phase space}
We take the molecular beam to be continuous (unbunched)  and
monoenergetic, in which case, it is completely specified by its density in the
($x, x', y, y'$) phase space (distribution function). The beam size in a
transport line depends, not on the density, but on the four-dimensional
volume ($V_{4D}$) occupied by the beam. This can be defined, experimentally, in a
number of ways: (a) the volume containing some fraction of the beam, (b) the
volume calculated from the root-mean-square (rms) beam sizes, or (c) the volume defined
by a set of collimators. These four dimensional volumes can be related to an equivalent theoretical
distribution function of Kapchinsky and Vladiminsky (KV)
\cite{kapchinsky59}. With the KV distribution function, we can calculate 
the linear focusing beam sizes ($a_x, a_y$) along the beam transport line 
and, in so doing, model the design optics. 

The KV distribution consists of a uniform density of particles on a 
hyper-ellipsoid in four-dimensional phase space. At a beam waist ($a ''_x = a ''_y =
0$), the hyper-ellipsoid is given by:
\begin{equation} \label{waist}
\left(\frac{x}{a_x}\right)^2 +\left (\frac{x'}{a_{x'}}\right)^2 +
\left(\frac{y}{a_y}\right)^2 + \left(\frac{y'}{a_{y'}}\right)^2  = 1 
\end{equation}
and has the volume, $V_{KV} = \frac{1}{2}\pi^2 a_x a_{x'} a_y a_{y'}$.
Typically we have beam waists at the beam source (minimum), at any
focus (minimum), and inside the focusing lenses (minimum or maximum).

The beam sizes along the beam transport line are given by the uncoupled
envelope equations, which for the equations of motion [Eq.~(\ref{linear_focus})] are:
\begin{eqnarray*} 
a''_x +K_0 a_x &=& \frac{\epsilon_{KVx}^2}{a_x^3} \nonumber\\
a''_y -K_0 a_y &=& \frac{\epsilon_{KVy}^2}{a_y^3} \label{envelope}
\end{eqnarray*}
where $ \epsilon_{KVx}$ and $\epsilon_{KVy}$ are the invariant
transverse emittances which, at a waist [see Eq.~(\ref{waist})], are
simply given by $ \epsilon_{KVx} = a_x a_{x'}$ and $ \epsilon_{KVy} = a_y
a_{y'}$.

If we project the KV distribution onto the ($x, x'$) plane, we obtain a
uniform density of particles inside an ellipse of constant area $\pi
\epsilon_{Kvx}$. For an arbitrary beam distribution, the equations
of motion,  for the rms beam sizes, have the same form as Eq.~(\ref{waist}). 
This defines the rms-equivalent KV distribution sizes as
$a_x = 2 \sigma_x$,  $a_{x'} = 2 \sigma_{x'}$, $a_y = 2 \sigma_y$,
and $a_{y'} = 2 \sigma_{y'}$. Under linear
forces,  a molecule remains on the same KV surface on which it started, 
with the shape of the ellipse changing but its volume remaining constant. 

Since the non-linear forces inside a lens become stronger, the further the molecule
is from the central axis, lenses may be evaluated
by computing the increase in non-linear effects with increasing beam size. 
For this we use nested KV distributions of increasing volume, each of which
is characterized by its volume, $V_{KV}$. The simulation results 
are then  independent of the characteristics of the initial beam and 
can be applied to non-surface beam distributions.
\subsubsection{beam survival\label{beam_survival}}
We calculate the survival of a 560 m/s methyl fluoride beam in the $J=0$ state, 
through a 30-m FODO lattice, as a function of the decapole strength ($a_5$). 
The lattice is described in section \ref{fodo}, and we assume that the beam is already
matched to the lattice. As molecular
beam jet sources are usually axi-symmetric, we take the emittance,
$\epsilon_{KV} $ to be the same in both transverse planes. Then, the initial
matched beam sizes and divergences, for the $\pi/3$ phase-advance
FODO cell, are:
\begin{eqnarray*}
a_{max}= a_x &=& (\epsilon_{KV} \beta_{max})^{1/2}    \nonumber \\   
a_{x'} &=& (\epsilon_{KV} \beta_{max})^{1/2} \nonumber\\
a_y &=& (\epsilon_{KV} \beta_{min})^{1/2}    \nonumber \\ 
a_{y'} &=& (\epsilon_{KV} \beta_{min})^{1/2} \label{matched_beam} 
\end{eqnarray*}
where $\beta_{min}= 0.587$ m  and $\beta_{max}=1.703$ m.
As $\beta$ is independent of the emittance, we use
the initial maximum beam size,
$a_{max}$ as our KV distribution size parameter :
\[V_{KV} = \frac{1}{2}\pi^2 \epsilon_{KV}^2 = \frac{\pi^2 a_{max}^4} {2 \beta^{2}_{max} } \]
The relation between the beam size and the emittance is shown in Fig.~\ref{KV}.
%
\begin{figure}
\includegraphics{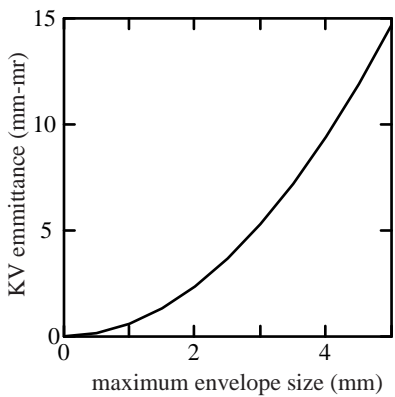}
\caption{\label{KV} Emittance versus beam envelope size of the shell in the
Kapchinsky-Vladiminsky  (KV) equilibrium beam distribution.}
\end {figure}
%

The trajectories were numerically integrated along
the FODO lattice using a $4^{th}$-order Runge-Kutta algorithm. 
We take as lost those particles whose
transverse displacement, in the beam transport line, in $x$ and/or $y$
became too large.  The transported particles, which started on a
zero-thickness KV-surface, finished up on a smeared-out fuzzy one, 
producing emittance growth and halo.

From the simulations, we found the fraction of the initial methyl fluoride beam surviving
as a function of position along the beamline. We did this for two different lens designs 
(two-wire field lens and optimized multipole) and for selected initial KV
beam sizes. The results are shown in FigÕs.~\ref{beam_loss} (a) and \ref{beam_loss} (b).

For the two-wire field lens, the largest KV beam that can be transmitted 
without loss is 1.4 mm. For the optimized multipole lens, the largest lossless
KV  beam is 2.5 mm. 
If the initial phase-space density of the beams is approximately constant,  the
relative beam intensity is given by the ratio of the lossless phase-
space volumes. \emph{This is a factor of} $(2.5/1.5)^4  \approx 10$
\emph{improvement  by using the optimized multipole
lenses} in place of the two-wire field lenses in this FODO lattice.

Figures~\ref{beam_loss} (a) and \ref{beam_loss} (b) 
also show that most of the beam losses occur in the first 5 m and by 30 m the 
losses are essentially complete. This suggests that extending the
beam transport line to much longer distances will not further reduce beam
survival. 
%
%
\begin{figure}
\includegraphics{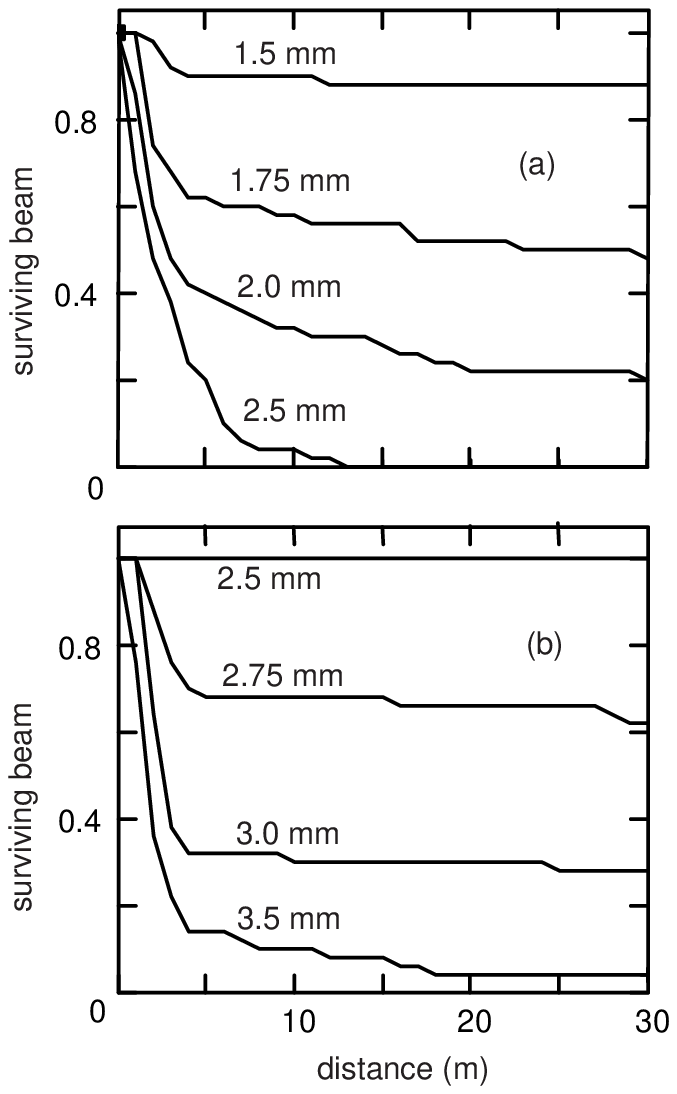}
\caption{\label{beam_loss} Calculated beam survival for 560 m/s methyl fluoride
in the $J = 0$ state in a 30 m-long FODO
lattice constructed of (a) two wire lenses or (b) optimized multipole
lens with $a_5 =0$. In each case $r_0$ = 6 mm and the central field, $E_0$ = 3.2 MV/m.
Survival is plotted as a function
of (KV) beam size.  For the 100\% beam survival, the
beam intensity scales as KV beam size to the fourth power (see text). 
Consequently, the FODO lattice of optimized multipole lenses will transport
about a factor of ten more beam than the same FODO lattice of two-wire field
lenses.
 }
\end {figure}
%
%

To find the most appropriate values of the lens decapole strength, 
for other molecules,
we also studied beam survival for the two limiting cases 
of strong electric field (linear Stark effect) and weak electric field (quadratic Stark effect).
A KV beam size of $a_{max} = 2.5$ mm was used and the results, along with the results
for methyl fluoride, are shown in Fig.~\ref{a5}.

For the strong field limit, as well as for the methyl fluoride example, we
obtain the highest transmission for $a_5 = 0$. In the weak electric field limit 
the optimum value of $a_5$ is shifted to $a_5 \approx -
0.2$, this nonlinearity in $\partial E/ \partial x$ compensating the
nonlinearity in $\partial W / \partial E$.
Thus, an $a_5$ near zero would be a good choice for 
CsF molecules in low-lying rotational states, 
and an $a_5$ near -0.2 would be a good choice for ground-state atoms. 
%
%
\begin{figure}
\includegraphics{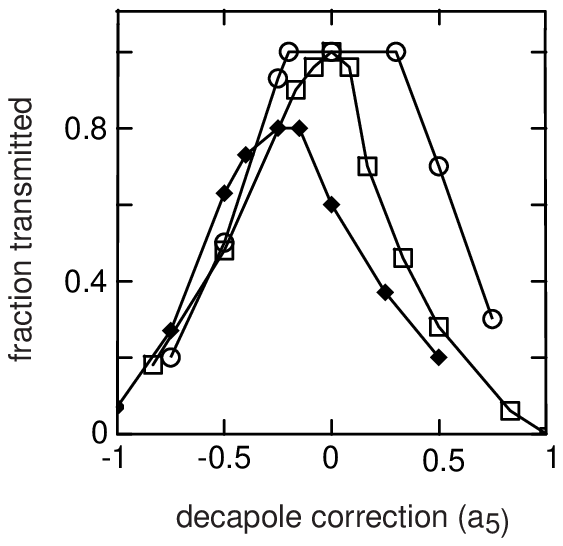}
\caption{\label{a5} Survival of 560 m/s monoenergetic beams through the 30
m FODO lattice for a KV beam with $a_{max} = 2.5 mm$ as a function of
decapole constant ($a_5$). Results for a methyl fluoride at 3.2 MV/m 
are shown as open squares,
the limiting cases of very strong and very weak electric field as 
open circles and filled diamonds respectively.
}
\end {figure}
%
%
\subsection {Velocity dependence of the beam
transmission\label{velocity_dependence}}
So far we have considered only a monoenergetic beam. 
To look at the tolerance of the different lenses to energy deviations,
we calculated the transmission for initially identical KV distributions
but with different energies (560 m/s being the matched velocity.)

The results are shown in Fig.~\ref{velocity}, for KV beam size of $ a_{max} = 1.5$ mm, 
for both the two-wire field lens and for the optimized multipole lens.
Both do quite well. At energies from 0.65 to 1.2 times the nominal energy
of  640 K (velocity 560 m/s),
transmission in the FODO lattice, of two-wire field lenses, is 50\% or more of 
its maximum value. The FODO lattice of optimized multipole lenses does 
even better, as we would expect, since it has lossless transmission up to a
KV size of $ a_{max}= 2.5$.
 One should note that Fig.~\ref{velocity} does
not represent the energy acceptance of a complete transport line, which will
be limited by the source and matching optics.
%
\begin{figure}
\includegraphics{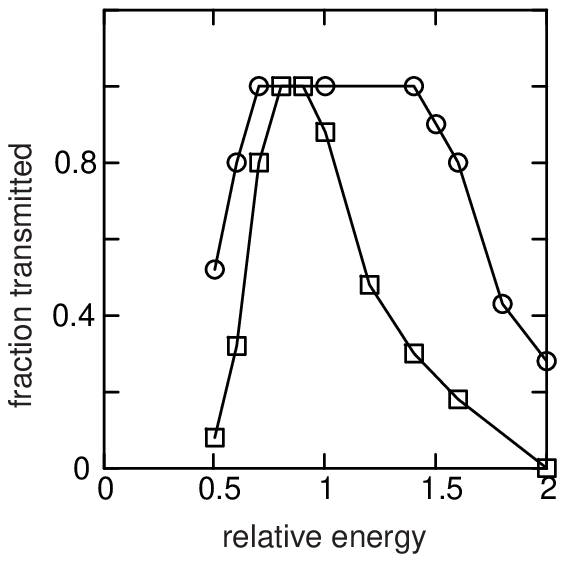}
\caption{\label{velocity} Calculated transmission, as a function of 
relative beam kinetic energy, of a 
beam of methyl fluoride in the $J = 0$ state through a 30-m FODO lattice.
The nominal energy is 640 K (560 m/s) and
the KV beam size is $ a_{max} = 1.5$ mm. Points for the FODO lattice using
two-wire field lenses and optimized multipole lenses (with decapole constant $a_5 = 0$)
are shown as squares and circles, respectively. }
\end {figure}
%
\subsection {Beam matching and point focusing\label{beam_matching}}
We complete the model transport line by adding matching optics 
upstream and downstream of the FODO lattice. This matches
the source to the lattice and the lattice to the final beam focus.
The matching, in both
cases, is achieved using a doublet- and a triplet- lens configuration as shown
in Fig.~\ref{beamline}. As the matching lenses require stronger
focusing, producing more non-linearities, we raise the lens
scaling length in the matching lenses to $r_0$ = 12 mm. And since most of the beam losses
occur early in the FODO lattice, we shorten the lattice to 15 m to reduce the
computing time necessary for the simulation. The beam envelope, in $x$
and $y$, and the placement of the lenses, is shown in Fig.~\ref{beam_envelope}, for an initial beam which
is close to the linear focusing limit in the lenses.
%
\begin{figure*}
\includegraphics{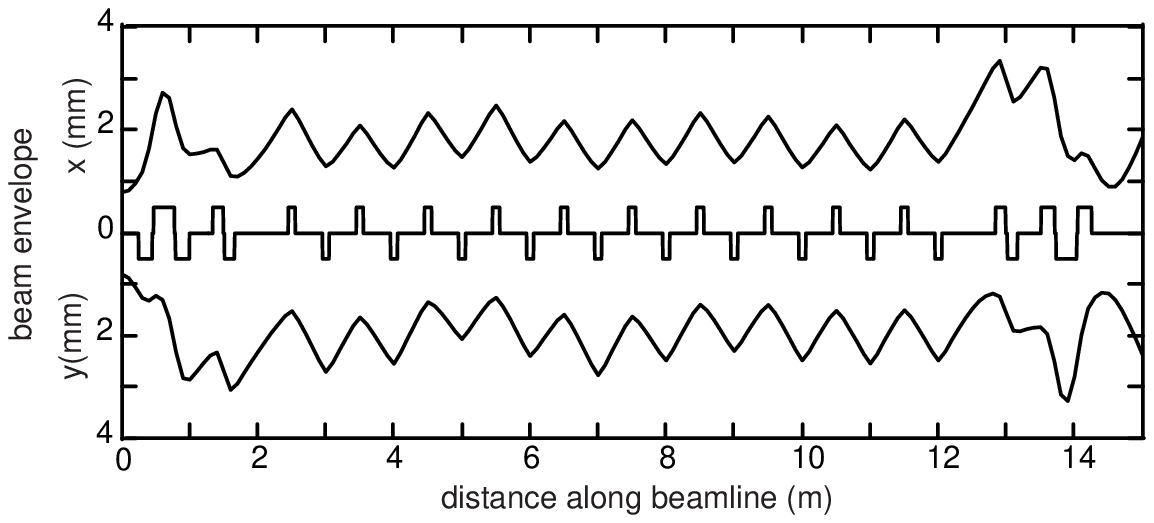}
\caption{\label{beam_envelope} Lens placement of a complete transport line
with matching lenses and a 15-m FODO lattice (center line), and the calculated rms beam
envelope, in $x$ and $y$, for 560 m/s beam of methyl fluoride in the $J = 0$ state.
We model the beam from a skimmed jet source and focus it to a 2 mm diameter spot.
A schematic of the beam transport line is shown in Fig.~\ref{beamline}. 
With a central field of 3.2 MV/m, scaling lengths of $r_0$ = 6 mm 
for the FODO lattice lenses, and 12 mm
for the matching lenses, the beam transport line has an acceptance of 2.5 mm
- mrad in both transverse planes. The longitudinal acceptance is about 84\% for a $\pm$ 10 \% rms energy spread.
}
\end {figure*}
%

For our example, of a 560 m/s beam of methyl fluoride in 
the $J = 0$ state, we assume that
the jet source has a very small orifice and a skimmer, of 1.5 mm diameter,
is placed 25 cm upstream of the first lens. An angular spread of $\pm$ 3.3
mrad is fixed by collimators (see Fig.~\ref{beamline}). 
This gives an initial beam emittance
of 2.5 mm-mrad in both transverse planes, and a maximum beam size of 2.1
mm in the FODO lattice. In addition, the initial beam is taken to have a
Gaussian energy spread of  $\pm$ 10 \% (rms). The calculated beam
transport, through this beamline, is 84\% of the entering
beam, the losses being due to the energy spread. 
Thus, we can transmit and focus most of the velocity distribution from the
jet source. (If needed, the beam can be focused, at the end, to an even
smaller size. However in this case, we have to increase the size of the
beam in the doublet/triplet lenses, where the non-linear forces will
produce emittance growth and beam halo. If we are not limited by the
electric field strength, we can correct this by using lenses with a larger
scaling length.)
\subsection {Transmitted intensity\label{intensity}}
If the characteristics of the initial molecular beam are known, 
the transverse and longitudinal acceptances 
of the full beam transport line may be used to calculate the beam intensity
at the final focus. As an example, consider the beamline in section \ref {beam_matching}, 
which has a skimmer radius of 0.75 mm and an angular acceptance of
$\pm$ 3.3 mrad.
From the point of view of an observer at the source, 
the entire beam transport line intercepts the same solid angle
as would a 2 mm diameter collimator, located 0.3 m from the skimmer, with
no lenses. The intensity of an unfocused beam at
this location can (often) be calculated or measured. 
From this, and a knowledge of the rotational state population fraction,
and the velocity spread, we can estimate the intensity at the final focus of the beamline.

For the methyl fluoride example, we assume a (seeded) jet source beam temperature of 3.5 K.
(This is based upon the equations in Ref.~\cite{miller88} for a 0.0035 cm 
diameter source orifice, 
a source pressure of $1.9 \times 10^5$ Pa (1400 torr) of Ar 
at a reservoir temperature of 300 K, and a methyl fluoride seed of 5\%. 
The source could be either pulsed or continuous.)
The 3.5 K, results in
a $J = 0$ population of about 30 \%, based upon a Maxwell Boltzmann
distribution,  and a kinetic energy spread
of 11.4\% which implies a longitudinal acceptance of about 60 \%. 

Thus an intensity, equivalent to 18 \% of a unfocused methyl fluoride beam passing through 
a 2 mm dia. collimator 0.3 m from the skimmer, would  reach 
the final focus some 14.5 m from the source.
And, since the beam transport line acceptances demonstrate  
transmission without loss, the same intensity 
should also reach a final focus much further from the source. 

Finally, we note that much higher performance beam transport lines can be
designed.  If one uses stronger electric fields, shorter FODO cells, fills more
of the beamline with focusing elements, 
and increases the size of the final focus, 
\emph{most of the solid angle from the skimmer can be accepted}.
Similar measures would allow one to efficiently transport and focus faster beams of molecules,
such as those seeded in a helium jet source.
\section{\label{uses}Applications}
Improving alternating gradient transport and focusing will make it easier to use
molecules in strong-field seeking states for experiments, for beam
transport, or to focus molecules for easier detection. 
In beam resonance experiments,
molecules prepared in a strong-field seeking state can be detected, 
after a transition to a weak-field seeking
state, as a Òflop-inÓ resonance. 

The optimized multipole lenses can greatly reduce the problem of Majorana
transitions\cite {majorana32, ramsey}. These are transitions that arise because different
$m_J$ levels belonging to the same $J$ are degenerate in zero field.
In very weak electric fields, a time-varying
component caused, for example, by the motion of the molecule through the
lens, can induce a transition to a different $ m_J$ state
with a very different Stark effect. This leads to beam loss, or
loss of signal and a large background in sensitive resonance experiments.

The optimized multipole lenses reduce this problem by allowing one to have
all normal or all skew F and D lenses with their central electric fields in the
same direction. This is done by changing the sign of $a_3$, the hexapole strength (and $a_5$)
while leaving the dipole strength ($a_1$) unchanged. 
Field direction changes in going between  F lenses and D
lenses are eliminated.  If a weak dipole (bias) field is added to the region
between the lenses, the molecules may never be in a rapidly changing
weak electric field.

The problem of Majorana transitions will be greatest for
molecules in weak-field seeking states focused by pure quadrupole and/or
sextupole lenses. These lenses have a vanishing field at the center and rapid
changes in the field direction at their entrances and exits \cite{reuss88}. 
The problem will be diminished if the molecules are focused in alternating gradient
fields using optimized multipole lenses with their central electric fields in
the same direction and
with alternating positive and negative values of $a_3$.  These lenses
have a non-vanishing field everywhere and can be optimized for a quadratic
Stark effect by choosing $|a_5| = -0.2$. 
Again a small dipole bias field can be used between lenses.

The problem of Majorana transitions can be eliminated by choosing the $J = 0$ state;
which is non-degenerate, always strong-field seeking, has the largest Stark effect of
any rotational level, and is highly populated in a cold jet source beam. 

The $J = 0$ and other strong-field seeking states  have unique and useful properties 
that can be exploited in experiments.
Within a rotational level, J, the $|m_J| = J$ states are strong-field seeking. 
States that are strong-field seeking in weak fields remain strong-field seeking 
in stronger fields
(but weak-field seeking states will become strong-field seeking in the limit
of strong fields).  Thus, for strong field seeking states, there is
no restriction on the size of the electric field that can be used to focus them. 
This is an advantage for molecules 
with small rotational constants and large dipole moments, where the weak
field-seeking states in low rotational levels become strong-field seeking in 
modest electric fields.
CsF (Fig.~\ref{stark}) and other heavy alkali halides are examples. 

Long distance beam transport, which can exceed 100 m or more, has a
number of applications. Since monatomic carrier gasses, clusters, and many contaminants will not
focus through the beam transport line, it can be used to clean up a beam. 
For hazardous and radioactive molecules, a long beamline allows one to separate
the source material and reservoir from the experiment and allows 
one to use radioactive detection in a lower background
environment. 

The long transit time (54 ms for the 30 m beam line in our example)
corrected, if necessary for the small longitudinal velocity changes in the
focusing elements, can be used for time-of-flight measurements 
with pulsed sources or a beam chopper. 
The different velocities will focus at slightly different longitudinal 
positions, which may be exploited
for position-sensitive detection. 
Alternatively, by using a pulsed beam and ramping the electric field in the (final
focus) lenses, all molecules may be brought to a focus, at 
different times, but at the same position. 

The long flight path may be useful for colinear laser excitation of weak
transitions. If the colinear laser excitation is combined with time-of-
flight measurement, the Doppler spread from the velocity distribution may
yield information about the absorption profile. Long transit time also
allows for the decay of some long-lived states. 
Molecular beams may be run in both directions to 
form a very long colliding beam apparatus. 
In this case, the reservoir temperatures of the beam
sources may be adjusted to equalize the focusing strengths of different molecules.

\begin{acknowledgments}
We thank Jason Maddi for performing the Stark effect calculation used in
Fig.~\ref{stark} and, along with Daniel Schwan, for early assistance with this
work. We thank Swapan Chattopadhyay for helping to get the work started.
This work was supported by the Director, Office
of Science, U.S. Department of
Energy, and, in part, by the Office of Basic Energy Sciences, U.S. Department
of Energy, both under Contract No. DE-AC03-76SF00098. 
\end{acknowledgments}

\bibliography{focusbib}

\begin{thebibliography}{20}
\expandafter\ifx\csname natexlab\endcsname\relax\def\natexlab#1{#1}\fi
\expandafter\ifx\csname bibnamefont\endcsname\relax
  \def\bibnamefont#1{#1}\fi
\expandafter\ifx\csname bibfnamefont\endcsname\relax
  \def\bibfnamefont#1{#1}\fi
\expandafter\ifx\csname citenamefont\endcsname\relax
  \def\citenamefont#1{#1}\fi
\expandafter\ifx\csname url\endcsname\relax
  \def\url#1{\texttt{#1}}\fi
\expandafter\ifx\csname urlprefix\endcsname\relax\def\urlprefix{URL }\fi
\providecommand{\bibinfo}[2]{#2}
\providecommand{\eprint}[2][]{\url{#2}}

\bibitem[{\citenamefont{Auerbach et~al.}(1966)\citenamefont{Auerbach, Bromberg,
  and Wharton}}]{auerbach66}
\bibinfo{author}{\bibfnamefont{D.}~\bibnamefont{Auerbach}},
  \bibinfo{author}{\bibfnamefont{E.}~\bibnamefont{Bromberg}}, \bibnamefont{and}
  \bibinfo{author}{\bibfnamefont{L.}~\bibnamefont{Wharton}},
  \bibinfo{journal}{J.\ Chem.\ Phys.} \textbf{\bibinfo{volume}{45}},
  \bibinfo{pages}{2160} (\bibinfo{year}{1966}).

\bibitem[{\citenamefont{Kakati and Lain{\'e}}(1967)}]{kakati67}
\bibinfo{author}{\bibfnamefont{D.}~\bibnamefont{Kakati}} \bibnamefont{and}
  \bibinfo{author}{\bibfnamefont{D.}~\bibnamefont{Lain{\'e}}},
  \bibinfo{journal}{Phys.\ Lett.} \textbf{\bibinfo{volume}{24A}},
  \bibinfo{pages}{676} (\bibinfo{year}{1967}).

\bibitem[{\citenamefont{Kakati and Lain{\'e}}(1969)}]{kakati69}
\bibinfo{author}{\bibfnamefont{D.}~\bibnamefont{Kakati}} \bibnamefont{and}
  \bibinfo{author}{\bibfnamefont{D.}~\bibnamefont{Lain{\'e}}},
  \bibinfo{journal}{Phys.\ Lett.} \textbf{\bibinfo{volume}{28A}},
  \bibinfo{pages}{786} (\bibinfo{year}{1969}).

\bibitem[{\citenamefont{Gunther and Sch{\"u}gerl}(1972)}]{gunther72}
\bibinfo{author}{\bibfnamefont{G.}~\bibnamefont{Gunther}} \bibnamefont{and}
  \bibinfo{author}{\bibfnamefont{K.}~\bibnamefont{Sch{\"u}gerl}},
  \bibinfo{journal}{Z.\ Phys.\ Chem} \textbf{\bibinfo{volume}{NF80}},
  \bibinfo{pages}{155} (\bibinfo{year}{1972}).

\bibitem[{\citenamefont{L{\"u}bbert et~al.}(1975)\citenamefont{L{\"u}bbert,
  G{\"u}nther, and Sch{\"u}gerl}}]{lubbert75}
\bibinfo{author}{\bibfnamefont{A.}~\bibnamefont{L{\"u}bbert}},
  \bibinfo{author}{\bibfnamefont{F.}~\bibnamefont{G{\"u}nther}},
  \bibnamefont{and}
  \bibinfo{author}{\bibfnamefont{K.}~\bibnamefont{Sch{\"u}gerl}},
  \bibinfo{journal}{Chem. Phy. Lett.} \textbf{\bibinfo{volume}{35}},
  \bibinfo{pages}{210} (\bibinfo{year}{1975}).

\bibitem[{\citenamefont{Lubbert et~al.}(1978)\citenamefont{Lubbert, Rotzoll,
  and Gunther}}]{lubbert78}
\bibinfo{author}{\bibfnamefont{A.}~\bibnamefont{Lubbert}},
  \bibinfo{author}{\bibfnamefont{G.}~\bibnamefont{Rotzoll}}, \bibnamefont{and}
  \bibinfo{author}{\bibfnamefont{G.}~\bibnamefont{Gunther}},
  \bibinfo{journal}{J. Chem. Phys.} \textbf{\bibinfo{volume}{69}},
  \bibinfo{pages}{5174} (\bibinfo{year}{1978}).

\bibitem[{\citenamefont{Reuss}(1988)}]{reuss88}
\bibinfo{author}{\bibfnamefont{J.}~\bibnamefont{Reuss}}, in
  \emph{\bibinfo{booktitle}{Atomic and Molecular Beam Methods}}, edited by
  \bibinfo{editor}{\bibfnamefont{G.}~\bibnamefont{Scoles}}
  (\bibinfo{publisher}{Oxford}, \bibinfo{address}{N.Y.}, \bibinfo{year}{1988}),
  p. \bibinfo{pages}{276}.

\bibitem[{\citenamefont{Noh et~al.}(2000)\citenamefont{Noh, Shimizu, and
  Shimizu}}]{noh00}
\bibinfo{author}{\bibfnamefont{H.-R.} \bibnamefont{Noh}},
  \bibinfo{author}{\bibfnamefont{K.}~\bibnamefont{Shimizu}}, \bibnamefont{and}
  \bibinfo{author}{\bibfnamefont{F.}~\bibnamefont{Shimizu}},
  \bibinfo{journal}{Phys. Rev. A} \textbf{\bibinfo{volume}{61}},
  \bibinfo{pages}{041601} (\bibinfo{year}{2000}).

\bibitem[{\citenamefont{Cho and Bernstein}(1991)}]{cho91}
\bibinfo{author}{\bibfnamefont{V.}~\bibnamefont{Cho}} \bibnamefont{and}
  \bibinfo{author}{\bibfnamefont{R.}~\bibnamefont{Bernstein}},
  \bibinfo{journal}{J. Phys. Chem.} \textbf{\bibinfo{volume}{95}},
  \bibinfo{pages}{8129} (\bibinfo{year}{1991}).

\bibitem[{\citenamefont{von Meyenn}(1970)}]{vonmeyen}
\bibinfo{author}{\bibfnamefont{K.}~\bibnamefont{von Meyenn}},
  \bibinfo{journal}{Z.\ Phys.\ A} \textbf{\bibinfo{volume}{231}},
  \bibinfo{pages}{154} (\bibinfo{year}{1970}).

\bibitem[{\citenamefont{C.H.Townes and Schawlow}(1955)}]{townes55}
\bibinfo{author}{\bibnamefont{C.H.Townes}} \bibnamefont{and}
  \bibinfo{author}{\bibfnamefont{A.}~\bibnamefont{Schawlow}},
  \emph{\bibinfo{title}{Microwave Spectroscopy}}
  (\bibinfo{publisher}{McGraw-Hill}, \bibinfo{address}{New York},
  \bibinfo{year}{1955}).

\bibitem[{\citenamefont{Schwan and Maddi}(2000)}]{schwan00}
\bibinfo{author}{\bibfnamefont{D.}~\bibnamefont{Schwan}} \bibnamefont{and}
  \bibinfo{author}{\bibfnamefont{J.}~\bibnamefont{Maddi}},
  \emph{\bibinfo{title}{private communication}} (\bibinfo{year}{2000}).

\bibitem[{\citenamefont{Miller}(1999)}]{miller99}
\bibinfo{author}{\bibfnamefont{T.}~\bibnamefont{Miller}}, in
  \emph{\bibinfo{booktitle}{CRC Handbook of Chemistry and Physics}}, edited by
  \bibinfo{editor}{\bibfnamefont{D.}~\bibnamefont{Lide}}
  (\bibinfo{publisher}{CRC}, \bibinfo{address}{Boca Raton},
  \bibinfo{year}{1999}), chap.~\bibinfo{chapter}{10}, p. \bibinfo{pages}{160},
  \bibinfo{edition}{79th} ed.

\bibitem[{\citenamefont{Miller and Bederson}(1977)}]{miller77}
\bibinfo{author}{\bibfnamefont{T.}~\bibnamefont{Miller}} \bibnamefont{and}
  \bibinfo{author}{\bibfnamefont{B.}~\bibnamefont{Bederson}}, in
  \emph{\bibinfo{booktitle}{Advances in Atomic and Molecular Physics}}, edited
  by \bibinfo{editor}{\bibfnamefont{D.}~\bibnamefont{Bates}} \bibnamefont{and}
  \bibinfo{editor}{\bibfnamefont{B.}~\bibnamefont{Bederson}}
  (\bibinfo{publisher}{Academic}, \bibinfo{address}{New York},
  \bibinfo{year}{1977}), vol.~\bibinfo{volume}{13}, p.~\bibinfo{pages}{1}.

\bibitem[{\citenamefont{Maddi et~al.}(1999)\citenamefont{Maddi, Dinneen, and
  Gould}}]{maddi99}
\bibinfo{author}{\bibfnamefont{J.~A.} \bibnamefont{Maddi}},
  \bibinfo{author}{\bibfnamefont{T.}~\bibnamefont{Dinneen}}, \bibnamefont{and}
  \bibinfo{author}{\bibfnamefont{H.}~\bibnamefont{Gould}},
  \bibinfo{journal}{Phys.\ Rev.\ A} \textbf{\bibinfo{volume}{60}},
  \bibinfo{pages}{3882} (\bibinfo{year}{1999}).

\bibitem[{\citenamefont{Bethlem et~al.}(1999)\citenamefont{Bethlem, Berden, and
  Meijer}}]{bethlem99}
\bibinfo{author}{\bibfnamefont{H.}~\bibnamefont{Bethlem}},
  \bibinfo{author}{\bibfnamefont{G.}~\bibnamefont{Berden}}, \bibnamefont{and}
  \bibinfo{author}{\bibfnamefont{G.}~\bibnamefont{Meijer}},
  \bibinfo{journal}{Phys. Rev. Lett.} \textbf{\bibinfo{volume}{83}},
  \bibinfo{pages}{1558} (\bibinfo{year}{1999}).

\bibitem[{\citenamefont{Kapchinsky and Vladimirsky}(1959)}]{kapchinsky59}
\bibinfo{author}{\bibfnamefont{M.}~\bibnamefont{Kapchinsky}} \bibnamefont{and}
  \bibinfo{author}{\bibfnamefont{V.~V.} \bibnamefont{Vladimirsky}}, in
  \emph{\bibinfo{booktitle}{Proc. Int. Conf. On High Energy Accelerators and
  Instrumentation}} (\bibinfo{publisher}{CERN}, \bibinfo{year}{1959}), p.
  \bibinfo{pages}{274}.

\bibitem[{\citenamefont{Miller}(1988)}]{miller88}
\bibinfo{author}{\bibfnamefont{D.}~\bibnamefont{Miller}}, in
  \emph{\bibinfo{booktitle}{Atomic and Molecular Beam Methods}}, edited by
  \bibinfo{editor}{\bibfnamefont{G.}~\bibnamefont{Scoles}}
  (\bibinfo{publisher}{Oxford}, \bibinfo{address}{N.Y.}, \bibinfo{year}{1988}),
  p.~\bibinfo{pages}{14}.

\bibitem[{\citenamefont{Majorana}(1932)}]{majorana32}
\bibinfo{author}{\bibfnamefont{E.}~\bibnamefont{Majorana}},
  \bibinfo{journal}{Nuovo Cimento} \textbf{\bibinfo{volume}{9}},
  \bibinfo{pages}{43} (\bibinfo{year}{1932}).

\bibitem[{\citenamefont{Ramsey}(1956)}]{ramsey}
\bibinfo{author}{\bibfnamefont{N.~F.} \bibnamefont{Ramsey}},
  \emph{\bibinfo{title}{Molecular Beams}} (\bibinfo{publisher}{Oxford},
  \bibinfo{address}{London}, \bibinfo{year}{1956}).

\end{thebibliography}
\end{document}